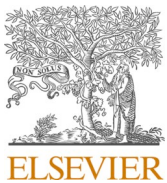
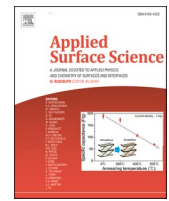

Full Length Article

# Growth interruption strategies for interface optimization in GaAsSb/GaAsN type-II superlattices

V. Braza [a], T. Ben [a], S. Flores [a,*], D.F. Reyes [a], A. Gallego-Carro [b], L. Stanojević [b], Ž. Gačević [b], N. Ruíz-Marín [a], J.M. Ulloa [b], D. González [a]

[a] *University Research Institute on Electron Microscopy & Materials, (IMEYMAT) Universidad de Cádiz, 11510 Puerto Real (Cádiz), Spain*
[b] *Institute for Optoelectronic Systems and Microtechnology (ISOM), Universidad Politécnica de Madrid, Avda. Complutense 30, 28040 Madrid, Spain*

A B S T R A C T

Recently, GaAsSb/GaAsN type II short-period superlattices (SLs) have been proposed as suitable structures to be implemented in the optimal design of monolithic multi-junction solar cells. However, due to strong surface Sb segregation, experimental Sb composition profiles differ greatly from the nominal square-wave design. In this work, the improvement of the interface quality of these SLs in terms of compositional abruptness and surface roughness has been evaluated by implementing different growth interruption times under $Sb_4/As_4$ (soaking) and $As_4$ (desorption) overpressure conditions before and after the growth of GaAsSb layers, respectively. The combined effects of both processes enhance Sb distribution, achieving squarer compositional profiles with reduced surface roughness interfaces. It has been found that the improvement in compositional abruptness is quantitatively much higher at the lower interface, during soaking, than at the upper interface during desorption. Conversely, a larger decrease in surface roughness is achieved at the upper interface than at the lower interface. Fitting of the Sb segregation profiles using the 3-layer kinetic fluid model has shown that the increase in Sb incorporation rate is due to the decrease in segregation energy, presumably to changes in the surface reconstruction of the floating layer at the surface.

## 1. Introduction

For several decades, multi-junction solar cells based on III-V semiconductors have remained the highest performing class of solar cells in terms of solar-electric conversion efficiency, approaching 40% efficiency: 37.9 % under the global AM1.5 spectrum [1]. However, its theoretical limit is still not reached due to the technological barriers to fabricating an ideal design with sub-cells with the right combination of lattice constants and bandgap energies. Therefore, the addition of a subcell with a suitable lattice constant and bandgap energy between 1.0 and 1.15 eV in the commercially available InGaP/GaAs/InGaAs solar cell structure could provide an optimal multilayer design to exceed an efficiency of 50% [2,3]. Recently, type II GaAsSb/GaAsN superlattices (SLs) have been proposed as a suitable structure to form such a 1.0–1.15 eV sub-cell, that would allow their implementation in the optimal design of GaAs lattice-matched multijunction solar cells [4,5]. Here, the authors suggest that this strategy optimizes control over the composition distribution of the different elements concerning bulk GaAsSbN counterparts, reducing the growing problems associated with concomitant Sb and N incorporation [6,7]. However, atomic-scale compositional analyses by electron microscopy in this type of SLs revealed that while the N distribution is spatially confined in the GaAsN layers, the same is not true for that of GaAsSb layers, there is a significant Sb segregation affecting both interfaces [8]. Thus, in the case of short-period SLs, the Sb composition profiles deviated strongly from the nominal square-shaped wave design adopting a shark-fin waveform in which the Sb content is never zero throughout the structure. In fact, it should be considered an SLs with a modulated composition made of GaAsSbN/GaAsSb, which also needs many periods until reaching the steady-state compositional condition [9]. Unfortunately, shorter period SLs are precisely the best ones to achieve an optimal trade-off between high photon absorption and high carrier extraction [10]. Indeed, interface quality is crucial for the optimization of device performance in terms of carrier confinement, carrier mobility, and photoresponse [11,12]. The way to approach the predicted photovoltaic properties in these SLs involves increasing the quality of the heteroepitaxial structure by improving the internal interfaces in terms of compositional abruptness and surface roughness at the atomic scale.

Numerous investigations have been conducted to achieve the best interfacial characteristics in layered heterostructures of III-V materials. Among them, III-V-Sb layers have attracted particular attention due to the enormous surface segregation of Sb and its effect on the interface quality. Several key factors have been shown to influence the interface attributes in these alloys, such as the growth temperature [13], the growth rate [14], nature of gas flux [15], surface morphology [16], or the surface strain [17]. To reduce the compositional gradient not only at the start of the Sb layer but also at the final interface, different growth approaches have been proposed. On the one hand, applying different growth interruption (GI) times with Sb exposure ("soaking") prior to the growth of a new Sb-containing layer has been associated with better structural quality and higher Sb incorporation [18–20], which leads to better optical performances [21,22]. In general, antimonides present weaker bonds compared to arsenides, so the As-for-Sb exchange is energetically favorable, leading to slower incorporation of Sb atoms into epitaxies with simultaneous As deposition [23]. However, the direction of this reaction could be reversed if we deliberately increase the Sb content at the surface. Better incorporation of Sb has been achieved in III-V layers that underwent pre-deposition or soaking steps with Sb and this result is related to changes in surface reconstruction. First-principles calculations using Density Functional Theory (DFT) suggest that Sb soaking promotes the formation of a primary surface layer richer in Sb-Sb dimers, which stimulates the substitution of As subsurface atoms incorporating Sb [24]. This agrees with several findings suggesting that at least the two topmost layers participate in the surface reconstruction, the structure of which differs from the bulk [25,26]. However, this enhancement is not endless, there is an optimal soak time for interfacial improvement in GaAs/GaAsSb interfaces that depends on the specific growth conditions. In fact, irregular interfaces in terms of surface roughness have been observed without soaking as well as long soak times (60 s), the best results obtained were with intermediate GI times (30 s) [27].

On the other hand, the enhancement of the upper GaAsSb/GaAs(N) interface by reducing the "floating layer" derived from Sb segregation both by rapid annealing processes [28] or after a GI with As flux exposure [18,19] has also been proposed. Annealing is not a preferable choice as it could significantly alter the composition and structure of buried layers by activating diffusion phenomena. The As-induced desorption process of Sb could be a better choice, but it should be applied carefully. Again, an optimal condition for As-soak time and temperature must be met [23] since long As-soak times could also harm the GaAsSb surface by causing large interfacial grading and generate large defect densities [29]. Certainly, longer GI times with As soaking contributes to higher out-diffusion of Sb at the exposed GaAsSb surface and it could lead to an important reduction of the Sb concentration below the top surface, which could be decisive in the case of very thin SL structures [30]. In addition, some authors pointed out that As soak processes only reduce interface roughness without improving the compositional gradient [31].

The present work aims to measure the effect of the GI processes on the interface quality at the nanometer level in GaAsSb/GaAsN SLs structures by using state-of-the-art scanning transmission electron microscopy (STEM). To this end, 5 different GI times were applied in conditions of $As_4/Sb_4$ (Sb soak) and $As_4$ overpressure (Sb desorption) before and after the GaAsSb layer growth, respectively. The improvements in the roughness and compositional gradients of both interfaces are evaluated and discussed within the kinetic segregation model framework.

## 2. Experimental procedure

One SL with 6 symmetric periods of GaAsN/GaAsSb was grown at different growth conditions by solid source molecular beam epitaxy (MBE) on GaAs (0 0 1) $n+$ substrates under $As_4$ overpressure conditions. Each period consists of 50 ML (~14 nm) of each alloy grown at a growth rate of 1 ML/s with different GI times. The layers have the same nominal flux for N (optical emission detection signal, OED, of 2.9 V) and Sb (beam equivalent pressure, $BEP_{Sb}$, of $7 \cdot 10^{-6}$ torr), corresponding to nominal contents of 2% and 5%, respectively. The sequence of growth of this SL is described in Fig. 1. First, after every GaAsN layer, an Sb soaking step was carried out by opening the Sb and As shutters to a fixed time (soak time, $t_1$) before the GaAsSb growth. Second, next to the GaAsSb layer deposition, the As shutter was opened during a fixed time (desorption time, $t_2$) before growing the next GaAsN layer. Six different GI times were used (0, 5, 10, 15, 20, and 25 s) this was the same time used for the corresponding soaking and desorption processes. Every period in the SL was called **P#** being # the used GI time.

Cross-sectional samples were prepared both by using the lamella method in a Focused Ion Beam microscope (FEI Scios™ 2 Dual Beam™) or by mechanical thinning followed by ion polishing. Then, the crystal quality of the structure was analyzed along the [1 1 0] direction by diffraction contrast (DC) TEM and annular dark field (ADF) in scanning TEM (STEM) mode while the Sb composition was quantified by Energy Dispersive X-Ray Spectroscopy (EDX) using ChemiSTEM technology with four embedded Bruker detectors. Talos F200X and FEI Titan Cubed³ Themis microscopes were used working at 200 kV.

## 3. Results

### 3.1. Effect in the interface abruptness

Fig. 2.a shows a representative DCTEM **g**002 DF image of the sample at a magnification of 250 kx. In well-established DCTEM conditions, **g**002 DF images are sensitive to the composition in zinc-blende structures since the intensity is proportional to the square of the structure factor as $4C(f_{III} - f_V)^2$ where $C$ is a factor that depends on TEM sample thickness and imaging conditions and $f$ is the atomic scattering factor for each element. Although this method is not suitable for quantitative analysis, the technique is sensitive for detecting small compositional changes in III-V heterostructures. In our case, for a $GaAs_{1-x-y}Sb_xN_y$ alloy where $x$ and $y$ are below 0.05, the presence of N in an atomic column reduces the image intensity while Sb increases it when compared to a 100% of As content of the substrate reference. Thus, relative to the substrate, the darker and brighter layers correspond to N-richer and Sb-richer regions, respectively. No period shows crystalline defects along the whole SL structure, indicative of a pseudomorphic growth, showing good periodicity and relatively flat interfaces. However, this perception can be misleading. Fig. 2.c plots the average intensity profile normalized with respect to the GaAs substrate along the growth direction. First, the interfaces are not so abrupt, especially in the first few periods. Secondly, it is observed that the intensities of the top GaAsN regions are always higher than the first GaAsN layer, which could be explained by the segregation of Sb in them. Unfortunately, the g002 imaging technique cannot resolve the combined contributions of Sb and N in the intensity

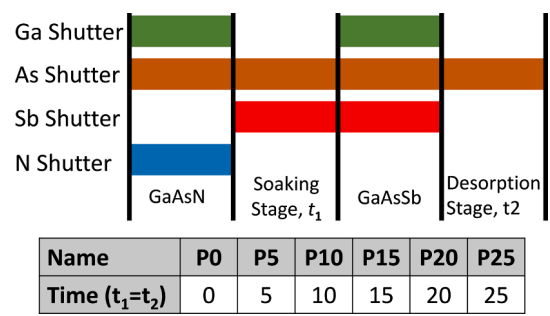

**Fig. 1.** Sketch of the shutter sequence of a period in the SL. The growth interruptions times and names for each period are noted in the table below.

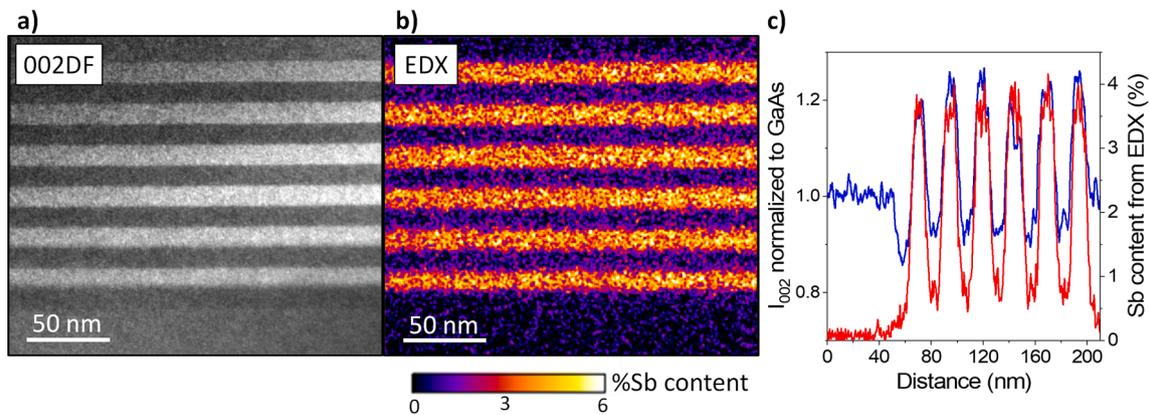

**Fig. 2.** General view of the sample using (a) DCTEM g002 DF imaging and (b) Sb mapping from EDX. (c) Profiles of the normalized intensity of the g002 DF image and of the average Sb content from the EDX mapping along the growth direction.

and the signal is also affected by the presence of deformation fields, which can significantly influence the contrast of the interfaces.

EDX analysis on cross-sectional TEM samples is a particularly useful tool to determine the distribution of Sb in SL structures, but not of N since it is not sensitive enough to study alloys with extremely low N contents. However, previous results had shown that the average N content is practically constant and independent of Sb segregation, being confined to the GaAsN [9]. Fig. 2.b shows the Sb distribution map obtained from EDX measurements of the same region. The elemental maps taken in different regions of the TEM sample are remarkably similar, ensuring high reproducibility of the measurements and growth homogeneity. As expected, the brightest layers in the DCTEM image coincide with the Sb-richer regions in the EDX map. To compare the compositional gradients at the interfaces, Fig. 2.c presents the linear profile of the Sb content obtained along the growth direction of the SL. The profile for the period without GI, **P0**, has a Gauss-like peak showing slow incorporation of Sb into the GaAsSb layer followed by a gradual decay towards the GaAsN layer. The other peaks progressively undergo a steepening effect as the GI time increases.

To characterize this effect with a higher resolution, EDX maps for each period were recorded at a magnification of 4.4 Mx. Fig. 3.a shows a superposition of all the profiles along the growth direction. As can be seen, the soaking and desorption processes produce significant differences in the shape of the profiles and important results can be extracted from the comparison of peaks. Firstly, the largest Sb content increases slightly from the period without GI (3.3 % Sb) to the longer GI time (3.8

% Sb). This increase is 12%, close to the 15% observed in other works [20]. More importantly, the plateau of stabilized Sb content is elongated. Since the maxima values are similar and it is difficult to assess the width of the plateau, the full width at half maximum (FWHM) of the different periods has been used as a measure of peak squareness (Fig. 3. b). The FWHM expands with the GI time, reaching a saturation state from the period **P25** onwards. As a result, the area under the profiles, which corresponds to the amount of Sb accumulated in each period, should increase in the same proportion, as can be seen in the same graph. Data extracted reveal that using the same time for the soaking and desorption processes implies net incorporation of Sb, with a gain of almost 30% when comparing the extreme cases. All of this affects the shape of the profiles, resulting in squarer layers with steeper interfaces.

Secondly, all these results are a consequence of increasing upward and downward gradients at both interfaces. Fig. 3.c represents the slope values at half peak height for the ascending and descending regions. It is easy to see that the change in slope steepness is notably larger in the upward region than in the downward region. The slope at the bottom interface improves by 250%, while at the top interface the slope increase is only 71%. The ratio between the upward and downward slopes, which compares the slope of each interface for a GaAsSb layer, shows a change, from a value of 0.56 with no GI (the upper interface is steeper) to 1.15 for the higher GI (the lower interface is better). The clear improvement in Sb incorporation is due to the reduction of the time to reach the composition plateau during Sb soaking. However, the increase in the slope seems to stabilize when the interruption time is 25 s. This could be the limit of Sb accumulation at the surface during the soaking stage, where longer soaking times would not imply an improvement of the interface. At the upper interface, stabilization is rather achieved after 15–20 s of desorption, which points to a shorter time for the desorption process to obtain the desired result. The combined effect of both processes (desorption and accumulation) contributes to the improvement of the profile area.

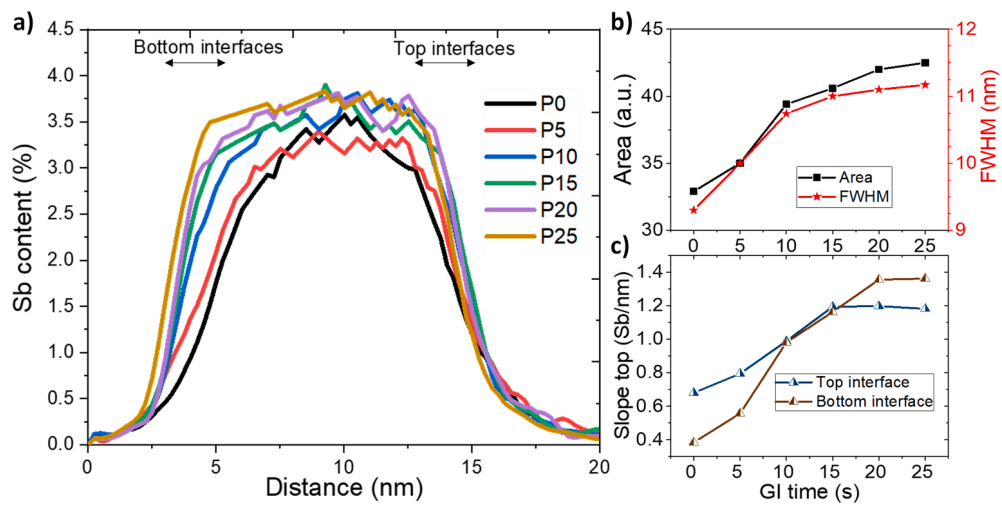

**Fig. 3.** (a) Superposed profiles of Sb content for each period extracted from high-magnification EDX mappings. (b) Area and FWHM calculated from every Sb profile. (c) Slopes obtained at half height of both interfaces in the Sb profiles (plotted as absolute values to more clear comparison).

### 3.2. Effect on the interface roughness

Many studies revealed that GI improves not only the compositional gradient but also decreases roughness [32–34]. To evaluate the flatness of the GaAsSb interfaces, the study was performed on an EDX map containing all layers acquired during one hour under stabilized conditions. The thickness of the sample hardly varies between the different zones. GaAsSb layers have been defined as the regions of the SL structure with an Sb content higher than 2%, which corresponds approximately to the FWHM of each layer. In Fig. 4.a, the boundaries of each GaAsSb layer are outlined as a yellow line in the EDX map of Sb content. Among roughness parameters, the root mean square ($R_q$) is the most widely used parameter for measuring the nanoscale roughness of the waviness height [35]. The $R_q$ is defined as:

$$R_q^m = \sqrt{\frac{1}{L}\int_0^L |Z_m^2(x)|dx} \qquad (1)$$

where $Z(x)$ is the function that describes the amplitudes respect to the mean value of the profile of the analyzed surface in terms of height (**h**) and position (*x*) of the sample over the evaluation length, *L*. Once the h(x) profiles are obtained, we calculate the mean and display it on the profile to observe that the mean is representative over the entire distance (zero slope). The values of Z(x) used in Eq. (1) are the difference between h(x) with respect to the mean value. The superscript *m* may refer to the upper interface (*up*), bottom interface (*bot*), or full-thickness (*th*).

It is necessary to clarify that our results are not intended to give the absolute values of the interface roughness, but a comparison of the roughness levels between the different periods. Certainly, the analyses performed with longer acquisition times show a decrease in the absolute values of the interface roughness. However, the trend of roughness improvement is maintained when comparing the periods with longer interruption times concerning the shorter ones. Taking all this into account, the roughness parameters of each period for the longest acquisition time are represented in Fig. 4.b. From comparison three outcomes can be drawn. First, all the parameters show a clear improvement above 5–10 s of GI times. Increasing the soaking and desorption times improves the interface roughness and reduces the thickness variation of the GaAsSb layers. Second, in all the periods, the bottom interfaces present lower $R_q$ values than the upper ones. Third, the improvement in the roughness by the GI is higher in the upper interfaces (34%) than in the lower ones (14%). Unsurprisingly, the desorption process is less effective to obtain abrupt interfaces compared to Sb soaking processes, but it achieves a higher improvement in the roughness reduction.

### 3.3. Modelling the segregation

Surface segregation in epitaxial growth has been investigated both theoretically and experimentally giving rise to several models of different complexity and origin, that may either be classified as phenomenological or atomistic [36]. Phenomenological models fit the experimental compositional data to profiles generated by geometrical series using effective segregation ratios [37,38]. Atomistic models simulate individual atomic site exchanges on the surface monolayers (MLs) that also can be grouped depending on the number of MLs involved. Such models employ systems of differential kinetic equations to simulate the temporal evolution of the concentrations during the epitaxial deposition of the two [39,40] or three [41,42] topmost superficial monolayers. A version of the latter, the 3-layers kinetic fluid model (3LKFM), has been historically applied to describe segregation in SiGe (group IV) alloys [43,44] but now it is widely used for exchange processes in group III–V alloys [9,45–47]. Comparative analyses point to 3LKFM as a refined option for describing segregation, at least in Sb alloys [45], due to the clear influence of the deeper layers in the anion exchange process, demonstrated experimentally [25] and theoretically [26]. It should be noted that the two-layers approach is limited as significant roughening can open paths for Sb segregation and 3LKFM may also take into account the role of atomic steps, kinks, and even surface reconstructions [48].

In this model, the two topmost layers closer to the surface are considered the floating layer in the exchange processes with respect to the bulk layer so the exchange between the three layers for every deposition time needs to be considered. The net exchange rate of As-for-Sb, $E_{i,i-1}$, to the layer *i* from the layer *i-1* is defined as the balance between the exchange process of Sb atom leaving the layer *i-1* and taking an As atom from the upper layer *i* in exchange and the reverse process:

$$E_{i,i-1} = P_1 X_{As}^i X_{Sb}^{i-1} - P_2 X_{As}^{i-1} X_{Sb}^i \qquad (2)$$

where $X_Y^i$ is the time-dependent concentration of element *Y* in layer *i*, and $P_{1,2}$ the exchange rate probabilities for each exchange. Each exchange process from a configuration $Sb_{i-1} + As_i$ to a $Sb_i + As_{i-1}$ is achieved by overcoming an energy barrier $E_1$ ($E_2$ for the reverse process) so the probability rates may be written as $P_{1,2} = \nu e^{-E_{1,2}/kT}$ where $\nu$ is the vibration frequency (~$10^{13}$ s$^{-1}$). $P_1$ and $P_2$ give the probabilities of surface segregation rates for As-for-Sb and Sb-for-As, respectively so they are connected for a given temperature. Thus, in the system of differential equations, the two exchange rate probabilities are dependent variables, so the parameter to fit with the experimental data is the ratio between them, $P_1/P_2$.

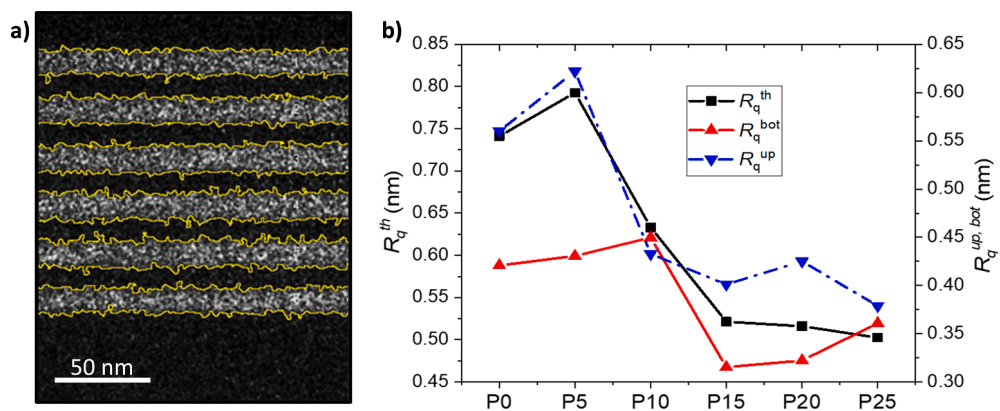

**Fig. 4.** (a) Boundaries of each GaAsSb layer are outlined as a yellow line in the EDX map of the Sb content, using a threshold of 2% of Sb. (b) Graph of $R_q$ values for the top interface (blue), the bottom interface (red) and the global thickness (black) for each period. (For interpretation of the references to colour in this figure legend, the reader is referred to the web version of this article.)

Fig. 5.a shows the simulation profiles with the best correlation to the experimental data for each period. We have assumed that the surface impingement rate, $\phi_{Sb}$, is constant during the growth in all periods and that the low content of N does not interfere with the Sb/As exchange mechanism. As can be seen, the simulated profiles match the experimental results, with a Pearson correlation coefficient higher than 0.96, although small mismatches are seen at the start and in the plateau of the profiles. Fig. 5.b shows the evolution of $P_1/P_2$ with the GI time, where the ratio decays from 3 to 2, which implies a change of 33%. It shows two stabilized regimes with a transition between 5 and 20 s of GI times.

## 4. Discussion

As we have seen, increasing both soaking and desorption GI times modifies the beginning and the end of the Sb segregation profiles during the GaAsSb layer growth, allowing for steeper interfaces with less roughness. It is broadly accepted that the cause of the surface segregation in III-V-Sb alloys is the formation of an Sb-richer floating layer over the epitaxial surface [18]. The Sb atoms accumulated in this floating layer are bound to the growth surface but do not contribute to the strain in the sample [49]. However, the type of surface reconstruction could change with an excess of Sb accumulation, as occurs during the soaking stage. Certainly, As-rich GaAsSb layers showed a clear c(4 × 4) reconstruction well known from GaAs(1 0 0) surfaces, whereas Sb-rich GaAsSb showed a (1x3) reconstruction, which was observed on GaSb(1 0 0) surfaces [34,50]. The accumulation of Sb results in the preferential formation of a structure of multiple Sb dimers on the surface due to the higher binding strength of Sb-Sb compared to Ga-Sb. [26,51]. The higher presence of these Sb dimers boosts the substitution of As atoms on the subsurface by Sb leading to more efficient incorporation of Sb [24] which would explain the decrease in the $P_1/P_2$ ratio during the Sb/As exchange. The $P_1/P_2$ ratio is function of the difference of the activation energies, $E_2-E_1$, usually known as the segregation energy, $E_s = E_2 - E_1 = kT\ln\frac{P_1}{P_2}$. As GI times are increased, the segregation energy decays from 0.071 eV to 0.037 eV due to an equalization of the exchange energy barriers $E_2$ and $E_1$. The opening of new pathways for substitution due to the formation of a dimer structure in the Sb-rich surface layers must be the cause. First-principles calculations performed in the framework of the DFT of the calculated energy differences of different substitution configurations are in this order [24].

At the top interface of the GaAsSb layer, the Sb accumulated in this floating layer at the end of GaAsSb growth should be gradually incorporated during the growth of the next layer, so floating layers with higher Sb contents should produce a greater increase in the width of the segregation profiles. The goal of the Sb desorption stage using As soaking is to reduce the Sb content in the floating layer by removing Sb atoms before the growth of the GaAsN layer. Two phenomena may occur to explain this behavior. First, Sb atoms can be desorbed from the floating layer and the top chemisorbed monolayer during growth interruption with an excess of As. For this, As atoms must be absorbed in the surface and then an As-for-Sb exchange occurs. Righi et al, by calculating the energetics of different As-for-Sb exchange configurations, found that the anion exchange process is energetically favored only when involving Sb atoms of the second Sb layer, rather than those belonging to surface ad-dimers [26]. Therefore, the dimer structure is maintained, and this explains why the segregation energy is the same throughout the process. Furthermore, this provides additional support for the use of a three-layer model to simulate the segregation process in GaAsSb layers. Secondly, the remaining Sb atoms in the growth chamber could be removed by the vacuum pump during the growth interruption, and thus fewer Sb atoms are incorporated into the structure [52]. However, the efficiency in altering the compositional profile is lower in the desorption stage than in the soaking stage, as we see by comparing the changes of the slopes. In any case, the improvement of this interface comes from the reduction of the roughness of the growth front, as other studies have observed [31,33]. Desorption seems to work mainly by reducing local inhomogeneities at the upper interface, resulting in a flatter segregation front. This effect should be more prominent in narrow quantum wells, leading to smaller perturbations of the electron confinement energy by reducing spatial fluctuations [53].

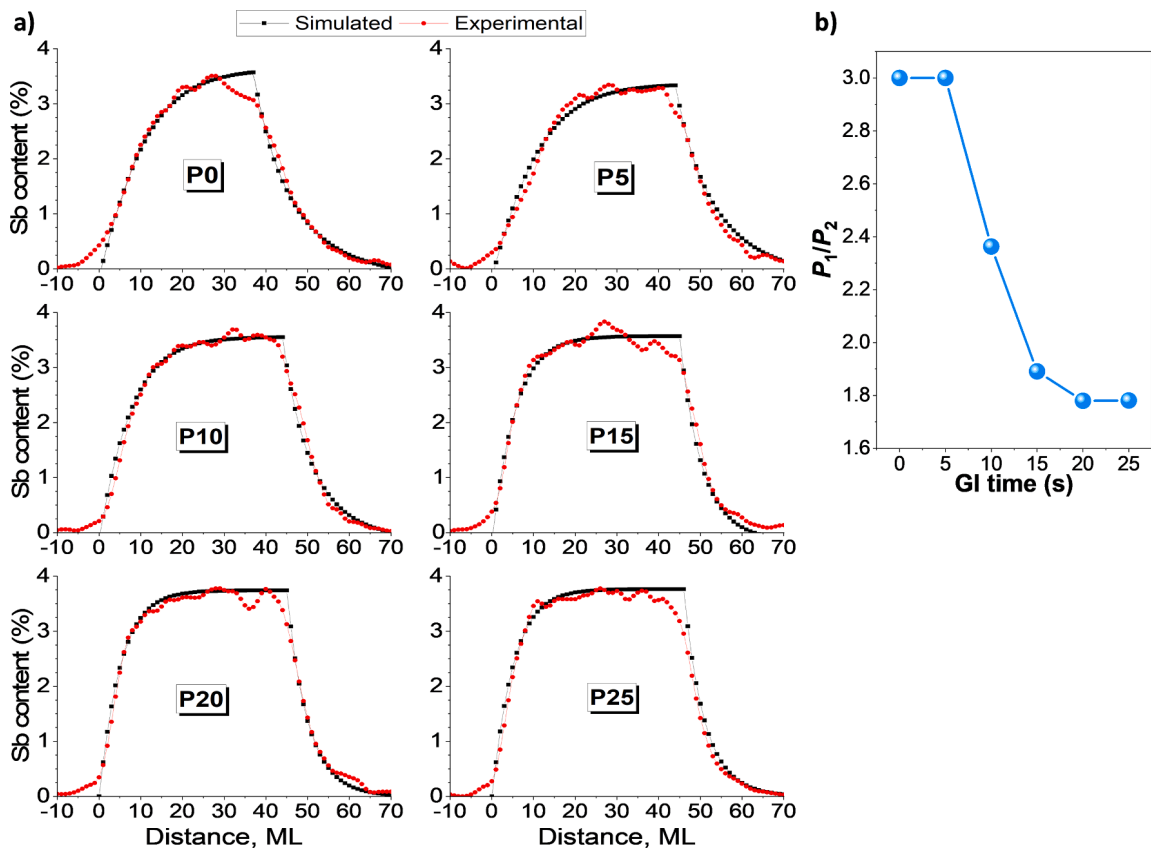

**Fig. 5.** (a) Experimental and simulated Sb profiles for every layer; (b) P1/P2 ratio vs interruption time.

Finally, the 3LKFM model has provided an excellent fit of the Sb profiles in the different periods of the SL, as expected from previous results [9]. However, the simulations performed with the model should be considered as a first approximation, since they do not consider the possibility that the surface reconstruction could be changing during growth, which likely occurs if the increase of Sb content in the floating layer does. Perhaps, one should consider the possibility that the exchange probability ratio varies during growth, which is taboo in all segregation models. Only Haxha *et al.* [45] have raised that possibility in an attempt to explain the slow increase in profiles at the onset of GaAsSb layer growth that we observe in all periods, and which is not described by any of the segregation models. The authors attribute it to changes in the strain energy during the deposition of the first monolayers, which would induce a penalty in the exchange energy. Whether variations in lattice mismatch or surface reconstruction are the reason for the changes in surface energy at the beginning of growth, this greatly affects the exchange probabilities, $P_1$ and $P_2$, as they have an exponential dependence on surface energy. In any case, this implies that the exchange probabilities may be different, at least initially, from those calculated during a steady state. Work is in progress to include all these aspects in the 3-layer fluid kinetic segregation model.

## 5. Conclusions

In this work, we have evaluated the effect of applying different growth interruption times (with Sb soaking at the beginning and Sb desorption at the end of each GaAsSb layer) on the quality of the interfaces in type II GaAsSb/GaAsN superlattices. Both processes have shown an improvement in both compositional abruptness and surface roughness, but to different degrees. Thus, the improvement of the slope is higher in the bottom interface during Sb soaking than in the upper during the desorption with As, while the opposite occurs for surface roughness. In both cases, a saturation of the enhancement effect is found for growth interruption times of about 30 s. Simulations of the Sb segregation using the 3LFKM model show a progressive decrease of the interchange probability $P_1/P_2$ ratio, driven by changes in the surface reconstruction as interruption times rise.

*CRediT authorship contribution statement*

**V. Braza:** Writing & editing – original draft, Investigation, Formal analysis. **T. Ben:** Draft review, Investigation, EM study, Project administration. **S. Flores:** Investigation, Writing – review & editing. **D.F. Reyes:** EDX Data curation, Writing – review & editing. **A. Gallego-Carro:** Investigation, Samples growth. **L. Stanojević:** Investigation. **Ž. Gačević:** Investigation. **N. Ruíz-Marín:** Investigation. **J.M. Ulloa:** Conceptualization, Project administration. **D. González:** Supervision, Methodology, Writing – review & editing, Project administration.

## Declaration of Competing Interest

The authors declare that they have no known competing financial interests or personal relationships that could have appeared to influence the work reported in this paper.

## Acknowledgments

The work has been co-financed by the Spanish National Research Agency (AEI project PID2019-106088RB-C33), the Regional Government of Andalusia (project FEDER-UCA18-108319), and the European Union's Horizon 2020 research and innovation program (Marie Skłodowska-Curie grant agreement n° 956548, project Quantimony). We would like to thank Francisco J. Delgado from the University of Cadiz for his experience in the preparation of FIB lamellas.